\documentclass{aa}
\usepackage{graphicx}
\usepackage{amsmath}
\usepackage{epsfig} 
\usepackage{txfonts,ulem}
\usepackage{xcolor}
\usepackage{url}
\usepackage{natbib}
\usepackage{longtable}
\bibpunct[]{(}{)}{;}{a}{}{,}

\newcommand{\avetilt}{$1.8\pm2.2^\circ$}   
\newcommand{\avexdist}{$20.3\pm0.6$~Mm}  
\newcommand{\aveydist}{$-1.1\pm0.7$~Mm}  

\newcommand{\avexdisttwo}{$44.7\pm1.4$~Mm}  

\newcommand{\slopedtiltgrey}{$-0.33 \pm 0.06^\circ$~per day}
 
\newcommand{\medianflux}{$4.6\times 10^{21}$~Mx}   

\newcommand*{\dt}[1]{\frac{d{#1}}{dt}}

\begin{document}

\title{
Average motion of emerging solar active region polarities \\II: Joy's law
}
\titlerunning{Emerging active regions: Joy's law}

\author{
H.~Schunker \inst{\ref{inst1},\ref{inst2}}
\and
C.~Baumgartner\inst{\ref{inst1}}
\and
A.~C.~Birch \inst{\ref{inst1}}
\and
R.~H.~Cameron\inst{\ref{inst1}}
\and
D.~C.~Braun  \inst{\ref{inst3}} 
\and
L.~Gizon \inst{\ref{inst1},\ref{inst4}}
}

\institute{
Max-Planck-Institut f\"{u}r Sonnensystemforschung, 37077  G\"{o}ttingen, Germany \label{inst1}\\
\email{schunker@mps.mpg.de}
\and
School of Mathematical and Physical Sciences, The University of Newcastle, New South Wales, Australia 
\label{inst2}
\and
NorthWest Research Associates, 3380 Mitchell Ln, Boulder, CO 80301, USA  
\label{inst3}
\and
Georg-August-Universit\"{a}t G\"{o}ttingen, Institut f\"{u}r Astrophysik, Friedrich-Hund-Platz 1, 37077   G\"{o}ttingen, Germany\label{inst4}
}

\date{Received $\langle$date$\rangle$ / Accepted $\langle$date$\rangle$}

\abstract
{
The tilt of solar active regions described by
Joy's law is essential for converting a toroidal field  to a poloidal field in Babcock-Leighton dynamo models. In thin flux tube models the Coriolis force causes what we observe as Joy's law, acting on east-west flows as they rise towards the surface.
}
{Our goal is to measure the evolution of the average tilt angle of hundreds of active regions as they emerge, so that we can constrain the origins of Joy's law.
}
{
We measured the tilt angle of the primary bipoles in 153 emerging active regions (EARs) in the Solar Dynamics Observatory  Helioseismic Emerging Active Region (SDO/HEAR) survey. We used line-of-sight magnetic field measurements averaged over 6~hours to define the polarities and measure the tilt angle up to four days after emergence.
}
{
We find that at the time of emergence the polarities are on average aligned east-west,  and that neither the separation nor the tilt depends on latitude. 
We do find, however, that EARs at higher latitudes have a faster north-south separation speed than those closer to the equator at the emergence time. 
After emergence, the tilt angle increases and Joy's law is evident about two days later.
The scatter in the tilt angle is independent of flux until about one day after emergence, when we find that higher-flux regions have a smaller scatter in tilt angle than lower-flux regions.
}
{
Our finding that active regions emerge with an east-west alignment is consistent with earlier observations, but is still surprising since thin flux tube models predict that tilt angles of rising flux tubes are generated below the surface.
Previously reported tilt angle relaxation of deeply anchored flux tubes can be largely explained by the change in east-west separation.
We conclude that Joy's law is caused by an inherent north-south separation speed present when the flux first reaches the surface, and that the scatter in the tilt angle is consistent with buffeting of the polarities by supergranulation.
}

\maketitle


\section{Introduction}\label{sect:intro}

There are two well-known constraints for dynamo models from early studies of flux emergence: Hale's law, which tells us that the magnetic bipoles of larger active regions that emerge in the northern and southern hemispheres have opposite polarities, and Joy's law, which describes the observed statistical tendency of the leading polarity of an active region to be closer to the equator than the following polarity \citep{Haleetal1919}.
This tilt angle between the leading and following polarities  tends to increase with  unsigned latitude \citep[e.g.][ and references therein]{vDGGreen2015}, and plays an essential role in the Babcock-Leighton dynamo model  \citep{Babcock1961,Cameronetal2015,KarakMiesch2017}.

Given the increase in tilt angle with latitude, the physical cause of Joy's law is believed to lie in the Coriolis force.
This immediately raises the question: Upon what motions does the Coriolis force act?  
One possibility is the motion associated with the buoyant rise of the magnetic flux tube through the solar convection zone \cite[e.g.][]{WangSheeley1991,DSilvaChoudhuri1993,FFH1995,Weberetal2011}. An alternative possibility is  the motion of the turbulent convection \cite[e.g.][]{Parker1955,ChoudhuriDSilva1990,Brandenburg2005}, with \citet{Schmidt1968} having suggested that active region bipoles emerge in upwelling supergranular cells with an east-west orientation, and that the surface flows in the cell move the polarities outwards, away from one another.  

Apart from the question of the motions involved in producing Joy's law, there is the question of what causes the large observed scatter in the tilt angle (Joy's law is a statistical tendency with large variations between individual active regions). This scatter has been found to be smaller for active regions with a higher flux \citep[e.g.][]{Jiangetal2014}.   
This could be due to the surface polarities with lower flux  being more susceptible to buffeting by the convection  (e.g.  \citealt{FFM1994,LongcopeFisher1996,Weberetal2011}),  or, as suggested in \citet{Schunkeretal2019},   the measurement of the position of larger polarities (with higher flux) has less scatter because the centre of gravity is not as affected by the buffeting by convection.

Traditionally, Joy's law has been  measured from continuum intensity images of sunspots \cite[e.g.][]{McClintockNorton2016}. The measured tilt angles are therefore of mostly well-established, stable active regions. To understand the origin of active region tilt angles it is therefore necessary to use magnetic field observations to capture the very beginnings of the emergence process, and to follow the evolution as a function of active region lifetime.

Observations from monitoring instruments such as  the  Michelson Doppler Imager on board the Solar and Heliospheric Observatory  \citep[SOHO/MDI; ][]{SOHO1995} and the Helioseismic and Magnetic Imager on board the Solar Dynamics Observatory \citep[SDO/HMI; ][]{SDO2012} make it possible to capture the emergence process of active regions both in intensity and magnetic field observations  \cite[e.g.][]{KosovichevStenflo2008,McClintockNorton2016}. 

\cite{KosovichevStenflo2008} studied more than 700 bipolar active regions using SOHO/MDI 96-minute cadence magnetic field observations, and found that the tilt angle of active regions at the time of emergence was statistically zero, and that the tilt angle is established during the emergence process (which lasts about 1\--1.5 days). In one case study using heliosesimology to measure subsurface flows,  \citet[][]{GHetal2013} showed that the direction of subsurface vortical flows below an anti-Joy's law active region (AR11073) is consistent with driving the leading polarity away from the equator. The advantage of using SDO/HMI is that it has observed hundreds of relatively simple active region emergence processes.

In this paper we present the statistical evolution of the tilt angle of 153 emerging active region (EAR) polarities from the Solar Dynamics Observatory Helioseismic Emerging Active Region (SDO/HEAR)  survey \citep{Schunkeretal2016,Schunkeretal2019} in an effort to understand the origins of Joy's law. 
Using  the SDO/HEAR  survey, \cite{Schunkeretal2019} identified two distinct phases of emergence. In phase~1   the speed of the separation between the polarities increases starting when the bipole first appears at the surface, and lasts until about $0.5$~days after the time of emergence. Phase~2 then begins when the speed of the separation starts to decrease and lasts until about two~days after the emergence time when the polarities stop separating. We   follow the evolution of the tilt angle in relation to these phases, and as a function of latitude to characterise Joy's law.
\citet{Birchetal2016} found by averaging over the emerging active regions in the SDO/HEAR  survey that there are no significant outflows during emergence, although these surface outflows are predicted by thin flux tube theory. \citet{Birchetal2019} did, however, find an average east-west elongated prograde flow just prior to emergence.
We now turn our attention to the evolution of the tilt angle and Joy's law.

First, in Sect.~\ref{sect:method}, we briefly describe how we measure the tilt angle of polarity pairs in emerging active regions from measurements in the SDO/HEAR  survey.
We then show the evolution of the tilt angle and the scatter in the tilt angle as a function of time and flux in Sects.~\ref{sect:tilttime} and \ref{sect:scat}.
In Sect.~\ref{sect:seplat} we show the north-south separation, east-west separation, and tilt angle as a function of latitude at the time of emergence and two days later. 
We discuss the change in tilt angle with latitude in relation to what we would expect from the Coriolis effect in Sect.~\ref{sect:discor}.
In Sect.~\ref{sect:relxn} we explain how the apparent tilt angle relaxation can be largely reproduced by the change in east-west separation.
We summarise our results in  Sect.~\ref{sect:summary} and discuss the models we think are useful to describe Joy's law.

\section{Measuring the tilt angle of the polarities}\label{sect:method}

We computed the tilt angle of the polarities in 153 active regions from  measurements of the location of the polarities as described in \citet{Schunkeretal2019}. The algorithm used in this paper was slightly modified from the previous measurements. We summarise the relevant details below.

The SDO/HEAR survey  \citep{Schunkeretal2016} consists of 182 emerging active regions observed by SDO/HMI between May 2010 and July 2014.
The $716 \times 716$~Mm ($512\times512$~pixel) Postel-projected maps of the SDO/HMI line-of-sight magnetic field are centred on the active region and tracked at the Carrington rotation rate up to seven days before and after the emergence.
We are interested in the evolution of the active regions on timescales of a fraction of a day.
For helioseismology purposes the data is divided into 6.825-hour   datacubes, and are labelled with a time interval (\texttt{TI}) relative to the emergence time ( \texttt{TI+00}).
We retained these time intervals for consistency and convenience and averaged the line-of-sight magnetogram maps over this time interval. Table~B.1 in \citet{Schunkeretal2019} lists the mid-time of the averaged \texttt{TI} to the time of emergence, $\tau=0$, for each time interval label.
They measured the position of the polarities in each of these averaged line-of-sight magnetic field maps using a feature recognition algorithm designed to determine the centroid position of the primary opposite polarities. 
The averaged line-of-sight magnetogram map was first shifted so that the centre of the map coincides with the centre of the active region  \citep[as defined by][]{Birchetal2013}. This was done using bilinear interpolation using the four nearest pixels which sometimes affected the identified location of the polarities, particularly in the more dispersed, following polarity at later times (see Appendix~\ref{app:noshift}). 
In this paper we first identify the locations of the polarities without shifting the maps, and then compute the locations relative to the active region centre. This procedure introduced differences in locations for some individual AR but this change does not affect the  previous results presented in \cite{Schunkeretal2019}.  Appendix~\ref{app:noshift} shows an example of the differences in position of the polarities for an individual active region and the resulting average position of the polarities.

Waves used for helioseismology are sensitive to scales larger than a few megametres, but to measure the location of the polarities it might be necessary to have a higher resolution. In Appendix~\ref{app:hires} we show that the  resolution of the maps does not significantly affect the average positions of the polarities or the tilt angle.

As in \citet{Schunkeretal2019} we excluded 29~active regions where it was difficult to track the locations of the polarities correctly, or where the active regions  have sustained anti-Hale orientation \citep[see Appendix~2 in][]{Schunkeretal2019}.  Our statistical analysis of the tilt angles was based on the remaining 153 EARs.

\begin{figure}
\includegraphics[width=0.45\textwidth]{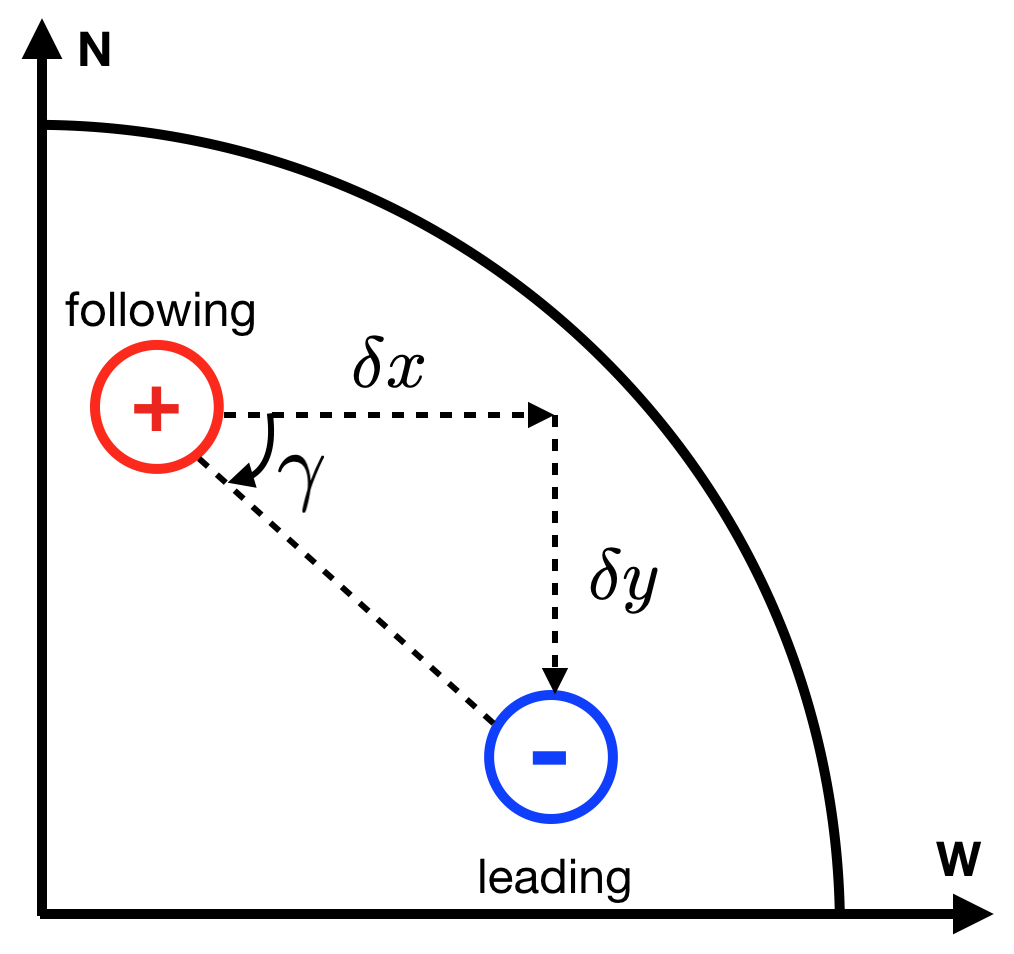}
\caption{Sketch of the orientation of a pair of polarities with a positive tilt angle in the northern hemisphere during solar cycle~24 where the leading negative polarity is closer to the equator than the following positive  polarity.}
\label{fig:tiltsketch}
\end{figure}

In \citet{Schunkeretal2019}, active regions in the southern hemisphere had their polarities inverted, so that they had a negative leading polarity and a positive following polarity as for northern hemisphere regions, and were flipped in the latitudinal direction to account for Joy's law.
Then the separation between the polarities in the $y$-direction, $\delta y (\tau) =  y_\mathrm{l} (\tau) - y_\mathrm{f} (\tau)$, is negative (positive) when the leading  polarity, $y_\mathrm{l}$, is closer  to (further from) the equator than the following  polarity, $y_\mathrm{f}$. 
The separation in the $x$-direction, $\delta x (\tau) =  x_\mathrm{l} (\tau) - x_\mathrm{f} (\tau)$, is defined as positive (negative) when the leading polarity is in the prograde (retrograde) direction from the following polarity.
From these measurements in \citet{Schunkeretal2019}, we define the tilt angle as
\begin{equation}
\gamma(\tau) = \arctan \left( \frac{- \delta y (\tau)}{\delta x (\tau)} \right).
\label{eqn:tiltdef}
\end{equation}
The tilt angle is positive when the leading  polarity is closer to the equator (and negative when it is further from the equator) than the following polarity (see Fig.~\ref{fig:tiltsketch}). 


\section{Tilt angle as a function of active region evolution}\label{sect:tilttime}

Understanding the origin and evolution of the tilt angle, as well as the dependence of the tilt angle on flux  will constrain models of active region emergence, and the location of the global toroidal magnetic field.

\begin{figure*}
\includegraphics[width=0.9\textwidth]{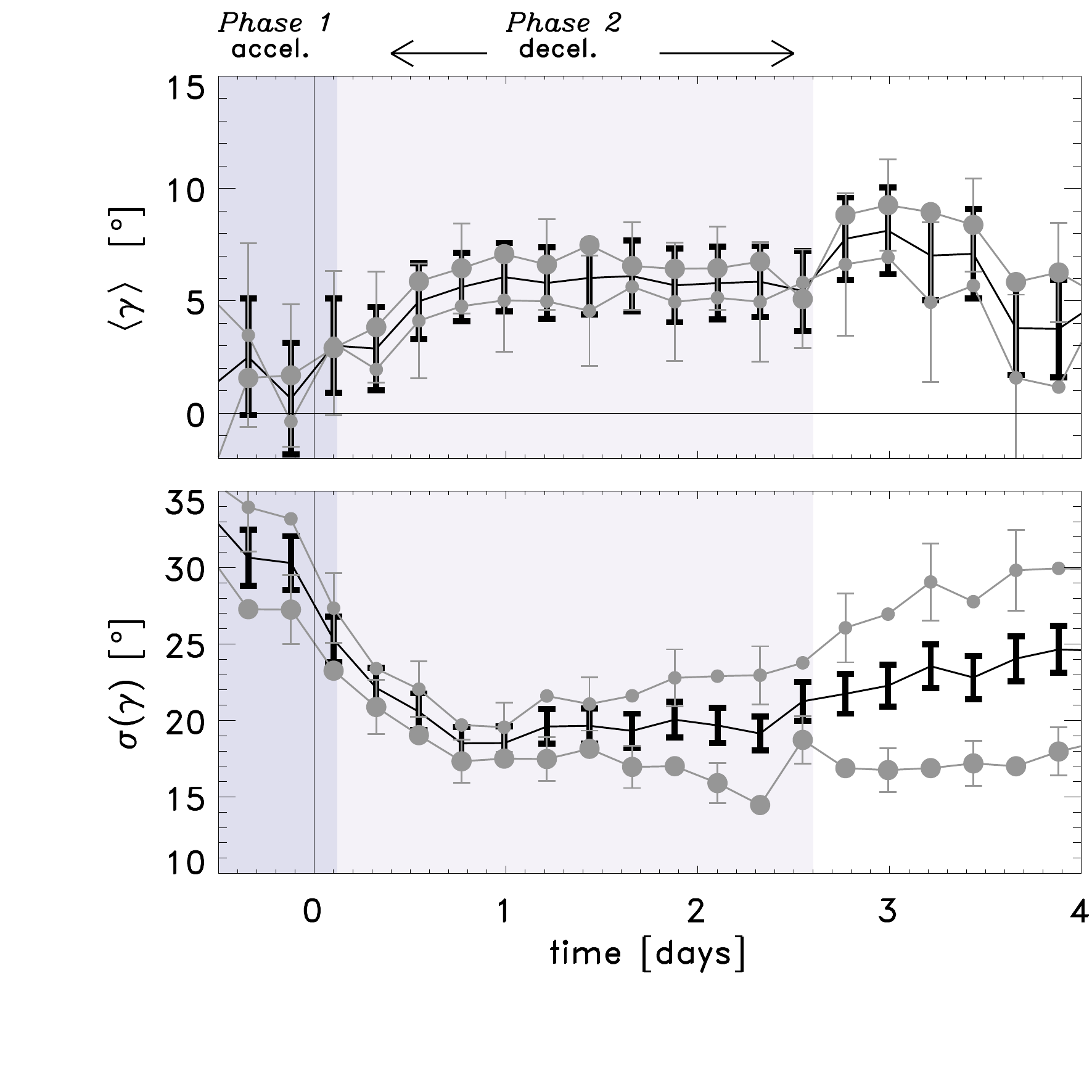}
\caption{Averaged tilt angle (top) and  standard deviation of the  tilt angle (bottom) of the polarities as a function of time for all EARs (black), EARs with a higher (lower) maximum flux than the median in large grey circles (small grey circles).
The EARs are divided into higher than or equal to, and lower than, the median maximum flux value,  \medianflux .
The standard deviation of the sample standard deviation at each time interval is described in \citet[][, Appendix~E]{Schunkeretal2019}. 
The shaded regions indicate two different phases of emergence, an increasing separation speed between the polarities followed by a decreasing separation speed \citep{Schunkeretal2019}.
}
\label{fig:tilttime}
\end{figure*}

We averaged the tilt angle over all valid active regions at each time step, as well as over regions with a maximum flux higher than or equal to, and lower than the median \medianflux .
In Table~A.1 of \cite{Schunkeretal2019}  active regions with a maximum flux value higher than or equal to the median of the 105 active regions used in \cite{Schunkeretal2016} have an asterisk. However, here we use the median value of the 153 valid active regions used in this paper.
We found that at the time of emergence the tilt angle is small, \avetilt (Fig.~\ref{fig:tilttime}, top panel)  and then increases over the course of the following day, after which it remains constant within the uncertainties. This excludes a constant tilt angle model, consistent with Fig.~2 in \citet{Schunkeretal2019}.
This figure also shows that there is no significant dependence of the tilt angle on flux.

\citet{FFH1995} found that at a fixed latitude, the tilt angle of white-light sunspot groups is smaller for polarities that are closer together, and hence have lower flux \citep{WangSheeley1989,Howard1992}.  
However, we have shown that, within the errors, the tilt angle (and north-south separation) does not depend on the eventual maximum flux of the active region, and only on the evolutionary stage of the active region:  large, high-flux  active regions also begin as small, low-flux active regions with negligible inclination.
Relative to our definition of emergence time \citep{Schunkeretal2016}, on average the active regions do not show unambiguous intensity darkening in the HMI full-disk continuum until about one day after emergence, and circular sunspots with a well-defined penumbra only form about two days after emergence. 
So another possible interpretation of the results in \citet{FFH1995} would be that many polarities that are close together have low flux and are near the beginning of emergence, whereas the polarities that are further apart have higher flux and are further evolved.

\section{Scatter of the tilt angle as a function of time and flux}\label{sect:scat}

We found that on average the tilt angle increases as an active region emerges, but the  evolution of the average tilt angle itself is not dependent on the maximum flux of the active region. Measuring the scatter in the tilt angle will help us to understand what causes the deviations from Joy's law. 

We found that the scatter in the tilt angle at the emergence time (when the active regions are small and close together) is large, $25 \pm 2^\circ$, and decreases, to about $20 \pm 1^\circ$, over the first day after emergence (bottom panel Fig.~\ref{fig:tilttime}). 
This is consistent with previous observations showing that there is less scatter in the tilt angle of high-flux regions \citep[e.g.][]{StenfloKosovichev2012,Jiangetal2014}, and demonstrates that the evolutionary stage of the active region is an important factor when characterising the tilt angle.

\citet{Schunkeretal2019} showed that the scatter in the motion of the polarities is largely independent of flux, but that the scatter increases with time and that the scatter of the leading polarity is systematically larger than the following polarity. The following polarity is known to be more diffuse than the leading polarity, and \citet{Schunkeretal2019} argued that both polarities are buffeted equally by supergranulation, but that the centre of gravity of the following polarity is not significantly affected by buffeting at its edges.
From this argument, larger, higher-flux regions would be expected to have less scatter in their tilt angles due to their larger size, and not a stronger resistance to buffeting by convection.
In the bottom panel of Fig.~\ref{fig:tilttime} we show that this is true after about one day after emergence. The scatter in the tilt angle for higher-flux regions remains roughly constant after this. However, the scatter in lower-flux regions continues to increase; this is probably due to the short lifetimes and decay of weak active regions \citep[e.g.][]{Schunkeretal2016}.

\section{Tilt angle and separation as a function of latitude}\label{sect:seplat}

Joy's law states that the average tilt angle of active region polarities increases with latitude, for example as measured by  \citet{WangSheeley1991} from line-of-sight magnetograms as $\sin \gamma = 0.48 \sin \lambda + 0.03$ where $\lambda$ is unsigned latitude.
To reflect this definition,  we multiplied the north-south displacement and the tilt angle by the sign of the latitude of the active region for the remaining analysis, $\mathrm{sgn}(\lambda) \gamma$ and $\mathrm{sgn}(\lambda) \delta y$. Therefore, active regions that obey Joy's law will have a negative tilt angle and positive north-south displacement in the southern hemisphere.

\begin{figure*}
\begin{center}
\hspace{-1cm}
\includegraphics[width=0.55\textwidth]{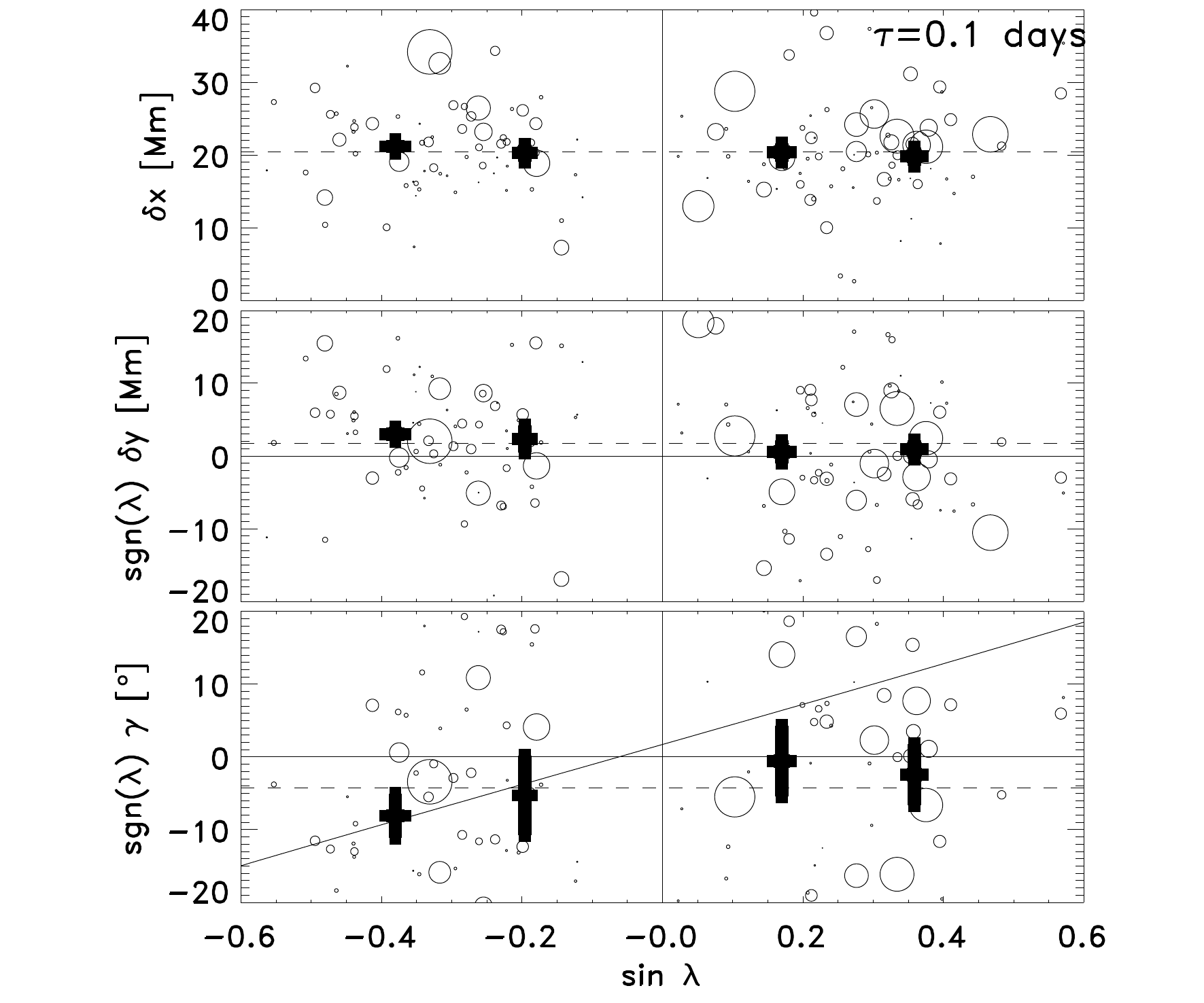}
\hspace{-1.5cm}
\includegraphics[width=0.55\textwidth]{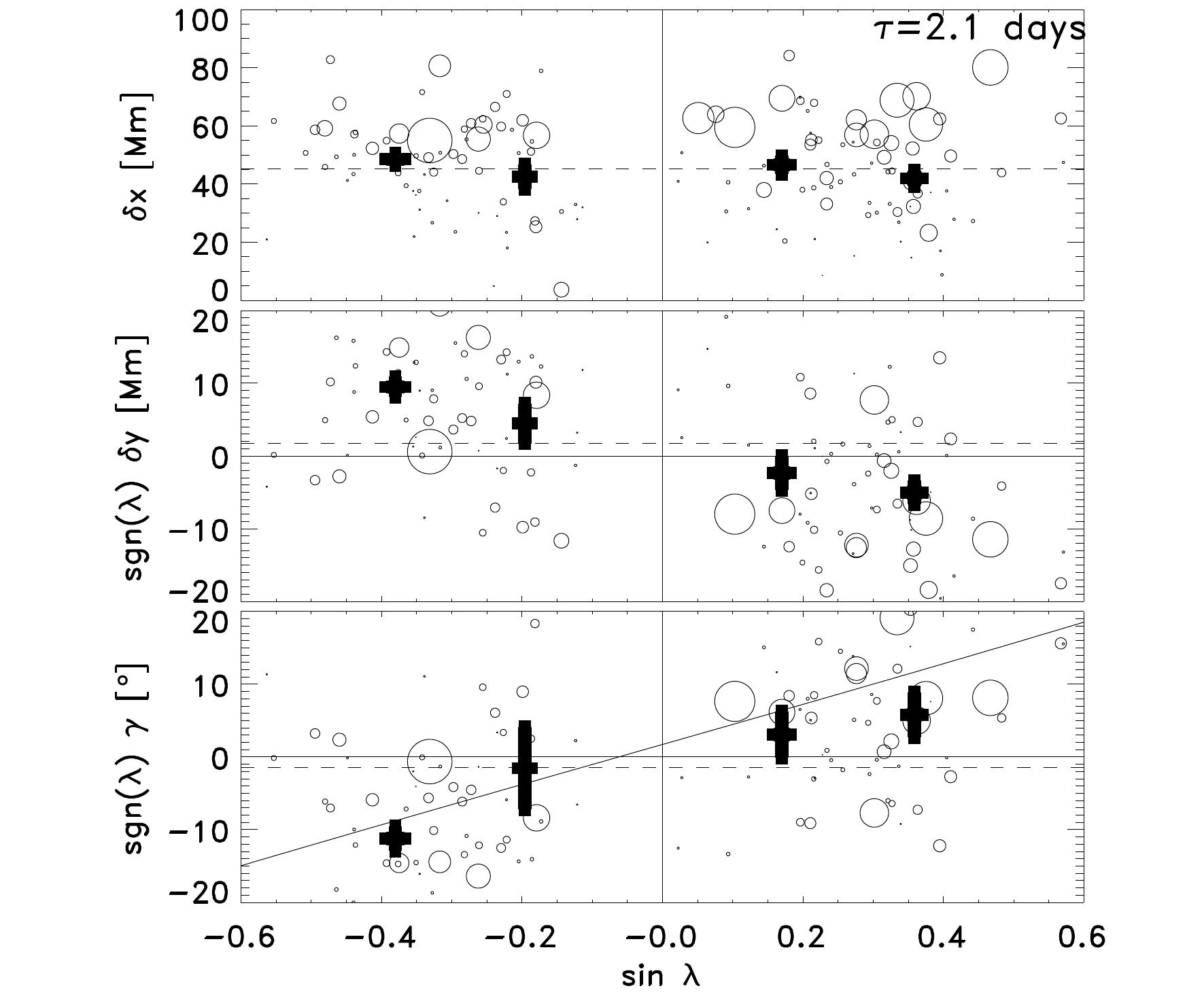}
\caption{
East-west separation, $\delta x$ (top); north-south separation, $\delta y$ (middle); and tilt angle, $\gamma$ (bottom) of the polarities as a function of latitude, $\lambda$, at the emergence time (left) and two days later (right).
The sign of the north-south separation and the tilt angle of active regions in the southern hemisphere have been adjusted,  i.e. in the southern (northern) hemisphere a negative (positive) $\delta y$ and a negative (positive) $\gamma$ is consistent with Joy's law \citep[black curve, ][]{WangSheeley1991}.
The size of the circle is proportional to the maximum flux of the active region.
The thick black points with error bars show latitudinal averages between 0 and 15$^\circ$ ($\sin \lambda = 0.26$) and 15$^\circ$ to 40$^\circ$ ($\sin \lambda = 0.64$) in the northern hemisphere, and the equivalent in the southern hemisphere.
The dashed lines are the mean values.
 }
\label{fig:deltasinelat}
\end{center}
\end{figure*}

We examined the longitudinal and latitudinal separation, as well as the tilt angle of the polarities as a function of latitude.  Figure~\ref{fig:deltasinelat} (left)  shows that at the emergence time, \texttt{TI+00}, the active regions have an east-west separation of about \avexdist, a north-south separation of \aveydist, \, and a small tilt angle of about \avetilt\ (recalling that there is already observable flux at the surface at this time). 
As expected from \cite{Schunkeretal2019},  low-flux regions tend to be closer together than large flux regions.
Joy's law is not evident because neither the separation nor the tilt angle  varies significantly from the mean. 
This is not consistent with thin flux tube simulations, where the flux tubes are tilted by the latitudinally dependent Coriolis effect acting as the tubes rise to the surface \cite[e.g.][]{Weberetal2011}. These simulations are valid in the regions where convection is relatively weak, and so in the remaining rise through the convection towards the surface this tilt angle would have to be somehow undone or hidden to accommodate the observations. 
However, the east-west orientation is consistent with the surface activity representing the subsurface toroidal flux \citep[e.g.][]{Parker1955,Cameronetal2017}.

In this section we have excluded ten additional active regions (11122, 11242, 11327, 11396, 11597, 11686, 11736, 11843, 11978, 12011) because they did not have valid position measurements at both $\texttt{TI+00}$ and $\texttt{TI+09}$.
There are three active regions that maintain a large anti-Joy tilt angle, 
two in the southern hemisphere AR~11400 ($\lambda=-14^\circ$, $\mathrm{sgn}(\lambda) \gamma = 80^\circ$ at $\texttt{TI+09}$), AR~11780 ($\lambda=-8^\circ$, $\mathrm{sgn}(\lambda) \gamma = 72^\circ$  at \texttt{TI+09}) and one in the northern hemisphere AR~11146 ($\lambda=23^\circ$, $\mathrm{sgn}(\lambda) \gamma = -68^\circ$  at \texttt{TI+09}). We  keep these active regions in our analysis; excluding these active regions does not change the results dramatically.
 
Figure~\ref{fig:deltasinelat} (right) shows the displacement and tilt angle of the active region polarities two days after emergence.
The east-west separation has increased to \avexdisttwo, retaining the expected flux dependence, and the north-south separation now varies with latitude, suggesting that whatever drives the north-south separation is responsible for the tilt angle. We find no dependence of the north-south separation on flux.

\section{Discussion of the Coriolis effect}\label{sect:discor}

The Coriolis force acts perpendicular to the direction of motion and to the axis of rotation. In the thin flux tube theory it acts on east-west flows in the flux tube driving a north-south displacement of the legs of the flux tube:  flux tubes with higher magnetic flux have faster east-west flows and  larger tilt angles. We do not find, however, any evidence of flux dependence in the tilt angle.

\citet{Schunkeretal2019} estimated the north-south separation speed numerically $\dt{\delta {y}}(i) = \left( \delta y(i+1) - \delta y(i-1) \right) / \left( \tau(i+1) - \tau(i-1) \right)$, where $i$ is the temporal index,  and similarly for the east-west separation speed $\dt{\delta {x}}$, and  $\dt\delta (i) = \sqrt{\dt{\delta {x}}(i)^2 + \dt{\delta {y}}(i)^2}$ . 
This revealed two clear phases of the emergence: Phase~1, when the speed of the separation between the polarities is increasing (accelerating), followed by Phase~2, when the speed is decreasing (decelerating). We indicated these phases for the tilt angle in Fig.~\ref{fig:tilttime}.

\begin{figure*}
\includegraphics[width=0.9\textwidth]{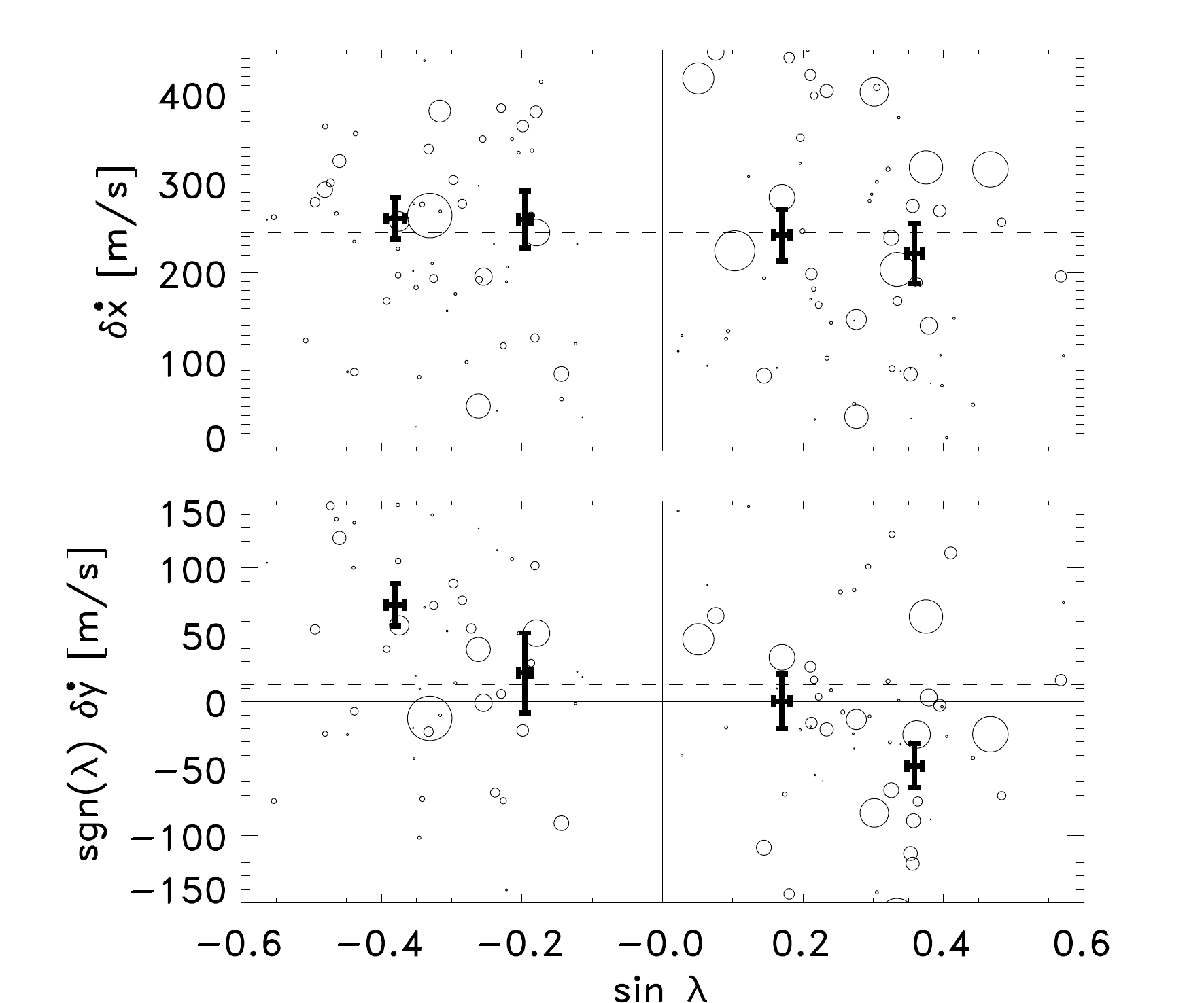}
\caption{ Separation velocity of the polarities in the east-west direction, $\dot{\delta x}$, (top panel) and the north-south direction, $\dot{\delta  y}$, (bottom panel) at the emergence time, \texttt{TI+00}, as a function of latitude, $\lambda$. 
The size of the circle is proportional to the maximum flux of the active region.
The thick black points with error bars show the averages over different ranges of latitude (between 0 and 15$^\circ$ ($\sin \lambda = 0.26$) and 15$^\circ$ to 40$^\circ$ ($\sin \lambda = 0.64$) in the northern hemisphere, and the equivalent in the southern hemisphere).
The dashed lines are the mean values.
 }
\label{fig:vxvysinelat}
\end{figure*}

The north-south separation speed is dependent on latitude at the time of emergence during Phase~1 (Fig.~\ref{fig:vxvysinelat}).
This demonstrates that the polarities emerge mostly east-west aligned, imbued with an inherent north-south velocity that is consistent with Joy's law.
It is not clear what drives this north-south velocity. Given its dependence on latitude, a natural conclusion is that the Coriolis force is responsible, but it is not clear upon which east-west velocities it is acting.
Naively, the Coriolis force should produce an acceleration in the north-south direction, but we see from Fig.~4 in \citet{Schunkeretal2019} that the acceleration of the separation 0.1~days after emergence, at the end of Phase~1, is zero within the uncertainties. This means that if the Coriolis force is acting, then it is only during a relatively short time to initiate the north-south velocity, or it is counteracted by an equal and opposite force. One candidate is the drag force \citep[e.g.][]{Fan2009}.
 
In Appendix~\ref{app:latsep} we model the expected separation of the polarities as a function of time and latitude for three models: a constant tilt angle model, the Coriolis effect acting on the east-west separation speed of the polarities, and a constant initial velocity model.
Our models show that a constant tilt angle model is not viable \citep[as already shown in Fig.~2, ][]{Schunkeretal2019}. However it is difficult to conclude anything further due to the large uncertainties in the separation of the polarities.

\section{Implications for the tilt angle relaxation}\label{sect:relxn}

\begin{figure*}
\includegraphics[width=0.9\textwidth]{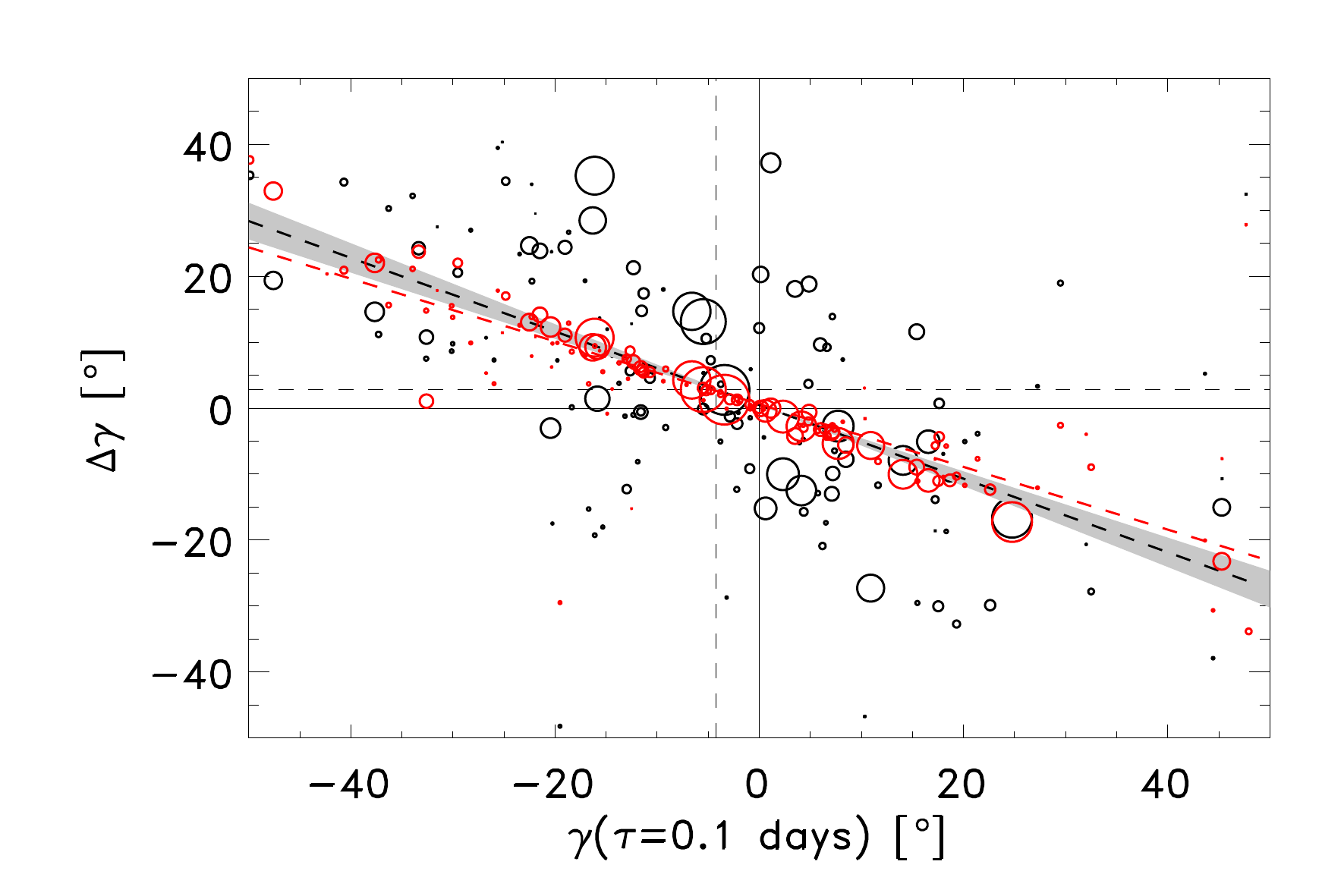} 
\caption{
Change in tilt angle, $\Delta \gamma$, between $\tau=2.1$~day and $\tau=0.1$~day as a function of the tilt angle, $\gamma$, at the emergence time (grey circles). 
The size of the circle represents the  maximum flux of the  active region.
The dotted grey line is a linear best fit to the observed $\Delta \gamma$ (grey circles) with a slope of $-0.65\pm0.06$ , and the shaded grey area indicates the uncertainty in the fitted slope parameter.
The red circles are the expected change in tilt angle for each EAR if  only $\delta x$ changed and $\delta y$ remained constant (see Eqn.~\ref{eqn:esttiltdef}). 
The red dotted  line is a linear best fit to the red circles with slope $-0.52\pm0.06$.
}
\label{fig:deltatiltdxcor}
\end{figure*}

\citet{Howard1996}  observed the tendency for the tilt angle  to move towards a more east-west orientation after emergence, which is not what is expected from the Coriolis force, and described it as a `relaxation'. 
This was interpreted in terms of magnetic tension by \cite{LongcopeChoudhuri2002}. In this interpretation the tilt angle evolves towards the position of the tube at the depth where the tube is disconnected, and they determined that this was likely to be occurring at the base of the convection zone. The initial scatter in the positions of the two polarities, imparted by the turbulent convective motions in the upper convection zone, should dissipate as the magnetic field at the surface becomes stronger and less susceptible to buffeting by the convective motions \citep[e.g.][]{LongcopeChoudhuri2002,TothGerlei2004}.  
In Fig.~\ref{fig:deltatiltdxcor} we also show that the tilt angles appear to develop a more east-west orientation, at a rate of \slopedtiltgrey \, (the change in tilt angle over two days).

\citet[e.g.][]{Schunkeretal2019} established that the average east-west separation of the polarities is larger than the average separation in the north-south direction. 
This east-west motion would cause a change in the measured tilt angle, rather than a circular motion of the polarities about a common centre.

To test this idea, we modelled the change in tilt angle, $\Delta \gamma_\mathrm{est}$, due to the change in the east-west separation of the polarities only, by using the measured $\Delta x = \delta x (\tau=2.1~\mathrm{days}) - \delta x (\tau=0.1~\mathrm{day})$ and leaving $\delta y$ constant at $\delta y (\tau=0.1~\mathrm{day})$. In Fig.~\ref{fig:deltatiltdxcor} the red circles represent
\begin{equation}
\Delta \gamma_\mathrm{est} = \arctan \left( \frac{- \delta y (\tau=0.1~\mathrm{day})}{\delta x (\tau=2.1~\mathrm{day})} \right) - \gamma (\tau=0.1~\mathrm{day}),
\label{eqn:esttiltdef}
\end{equation}
and we can see that this reproduces much of the apparent relaxation.

If we subtract the model tilt angle, $\Delta \gamma_\mathrm{est}$, the dependency of the change in tilt angle on the initial tilt angle vanishes.    
From our analysis of the independent motion of the polarities we have demonstrated that what was previously   interpreted as a tilt angle relaxation is a straightforward consequence of the east-west separation of the polarities.
Any constraints placed on models of emerging flux tubes using the apparent tilt angle relaxation need to be carefully reconsidered.

\section{Summary and discussion}\label{sect:summary}

Our finding that, on average, active regions emerge  with an east-west alignment is consistent with earlier observations, but is still surprising since thin-flux-tube models predict that tilt angles of rising flux tubes are generated below the surface.

Our results show that the forces driving Joy's law are observed as an inherent north-south separation speed of the polarities that depends on latitude but is independent of flux.
The origin of the north-south separation remains unclear.
Our results indicate that if it is due to the Coriolis effect acting on flows within the emerging flux tube, then the flows in the tube must be largely directed away from the loop apex and independent of flux.

\citet{Chenetal2017} simulate the emergence of a thin flux tube through the top 20~Mm of the convection zone. The locations of the polarities at the surface lie above the location of the polarities at the footpoints (bottom of the box). The simulations do not include solar rotation per se, but the time evolution of the flux tube at the bottom boundary does. The simulation of one single active region cannot be directly compared to an average of many active regions, and so we are hesitant to compare the tilt angle development.

One explanation for the initial observed east-west orientation is that the initial emerging flux tube has the correct amount of twist and writhe \cite[e.g.][]{LDMPvDG2003} so that the field at the apex of the emerging loop is east-west aligned. When the apex breaks the surface,  the twisted field is aligned east-west, with Joy's law developing as the writhe becomes more evident.

It is known that there is a relationship between the supergranulation pattern and where flux emergence occurs \citep{Birchetal2019}. We speculate that if the supergranulation is guiding the initial emergence process an alternative explanation for our results is that the emergence into predominantly east-west aligned north-south converging flows \citep{Birchetal2019} leads to a preference for east-west alignment of the polarities.
Why the emergence location is preferentially in east-west aligned inflows is not clear.

Our findings are consistent with the model of emerging flux as presented in \citet{Schunkeretal2019}. During Phase~1, active region polarities emerge east-west aligned (zero tilt angle) with an increasing separation speed, which lasts until about 0.5~day after the emergence time, and the tilt angle begins to develop.  
Phase~2 begins when the separation speed starts to decrease, until the polarities stop separating about $2.5\--\,3$~days after the time of emergence. The latitudinal dependence of the tilt angle, characteristic of Joy's law sets in during this second phase. In the first day after emergence, the scatter in the tilt angle decreases independent to the maximum flux, consistent with the polarities being buffeted by near-surface convection as they move to lie over their footpoints anchored at some depth below the surface.  
Analysis of the flows at and below the surface leading up to the emergence will help to constrain the subsurface picture.

\begin{acknowledgements}
CB  is a member of the International Max Planck Research School for Solar System Science at the University of G{\"o}ttingen. 
CB tested the reproducibility of the results and high-resolution tests, specifically Appendix~A and Appendix~B.  
The German Data Center for SDO, funded by the German Aerospace Center (DLR), provided the IT infrastructure for this project.  
Observations courtesy of NASA/SDO and the HMI science teams. 
ACB, RHC, and LG acknowledge partial support from the European Research Council Synergy Grant WHOLE SUN \#810218.
This work utilised the Pegasus workflow management system. 
DCB is supported by the Solar Terrestrial program of the US National Science Foundation (grant AGS-1623844) and by the Heliophysics Division of NASA (grants 80NSSC18K0068 and 80NSSC18K0066).
\end{acknowledgements}

\bibliographystyle{aa} 
\bibliography{ears.bib} 

\appendix

\clearpage 


\clearpage
\section{Polarity centres identified by shifting the search mask for AR~11072}\label{app:noshift}

As described in Section~2 of \cite{Schunkeretal2019}, we define a search area to be limited to all pixels within a radius of 100~Mm from the centre of the map, with magnetic field strength averaged over all active regions greater than 10~G. This resulted in a roughly circular search area at the centre of the map for early time intervals, which increased in size, and became more elliptical with the semi-major axis in the east-west direction in time. The search mask is the corresponding map where pixels that satisfy this condition have a value of 1, and 0 otherwise.

 We then shifted the individual line of sight magnetic field maps so that the location of emergence was at the centre. This shift used a bilinear interpolation  over four pixels. We found that at later times, when the following polarity is more dispersed, if we did not shift the maps, but instead shifted the search mask to lie over the location of the emergence, the displacement of some of the features relative to the active region centres was significantly different.

\begin{figure*}
\includegraphics[width=0.9\textwidth]{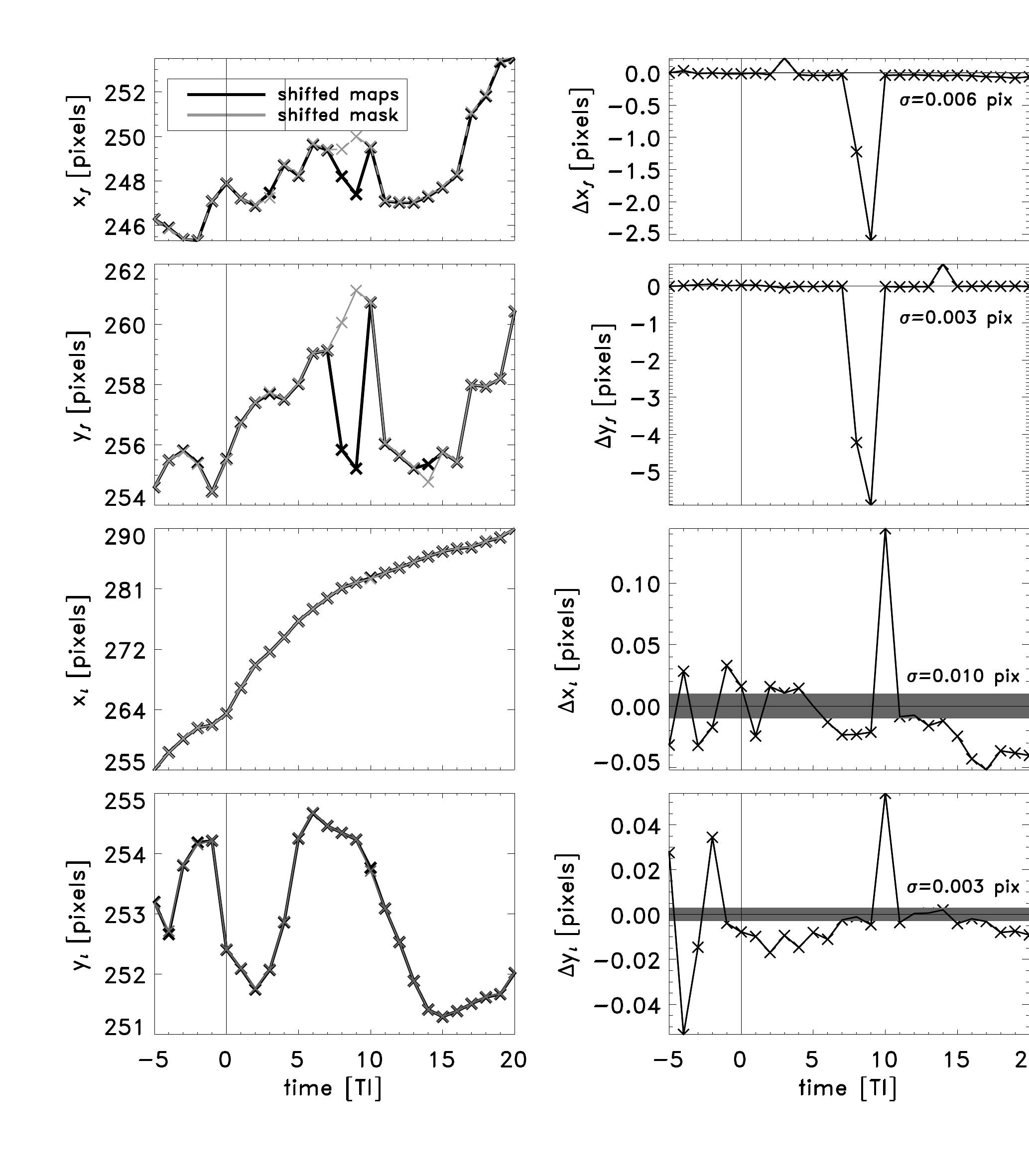}
\caption{
Comparison of polarity centres identified by shifting the magnetic field maps (previous algorithm) and shifting the search mask (updated algorithm) for AR~11072. The left column shows the  feature locations from the previous algorithm (black) and the updated algorithm (grey) in the $x$ and $y$ directions of the following polarity ($x_f$, $y_f$, top two rows) and leading polarity ($x_l$, $y_l$, bottom two rows). The right column shows the difference between the previous and updated algorithm. The grey shaded regions indicate the uncertainty in the data (also given by $\sigma$) from averaging every fourth image of  \texttt{TI+02} datacube of AR~11072 and finding the feature  \citep[as shown for AR~11075 in ][, Appendix~D]{Schunkeretal2019}. The differences can be outside of the uncertainties in the data.
 AR~11072 is an active region in the southern hemisphere. The active region is shown with the sign of the polarities inverted and the map reversed in the latitudinal direction, as  used in the statistical analysis. 
}
\label{fig:posalg2}
\end{figure*}

\begin{figure*}
\vspace{1cm}
\includegraphics[width=0.9\textwidth]{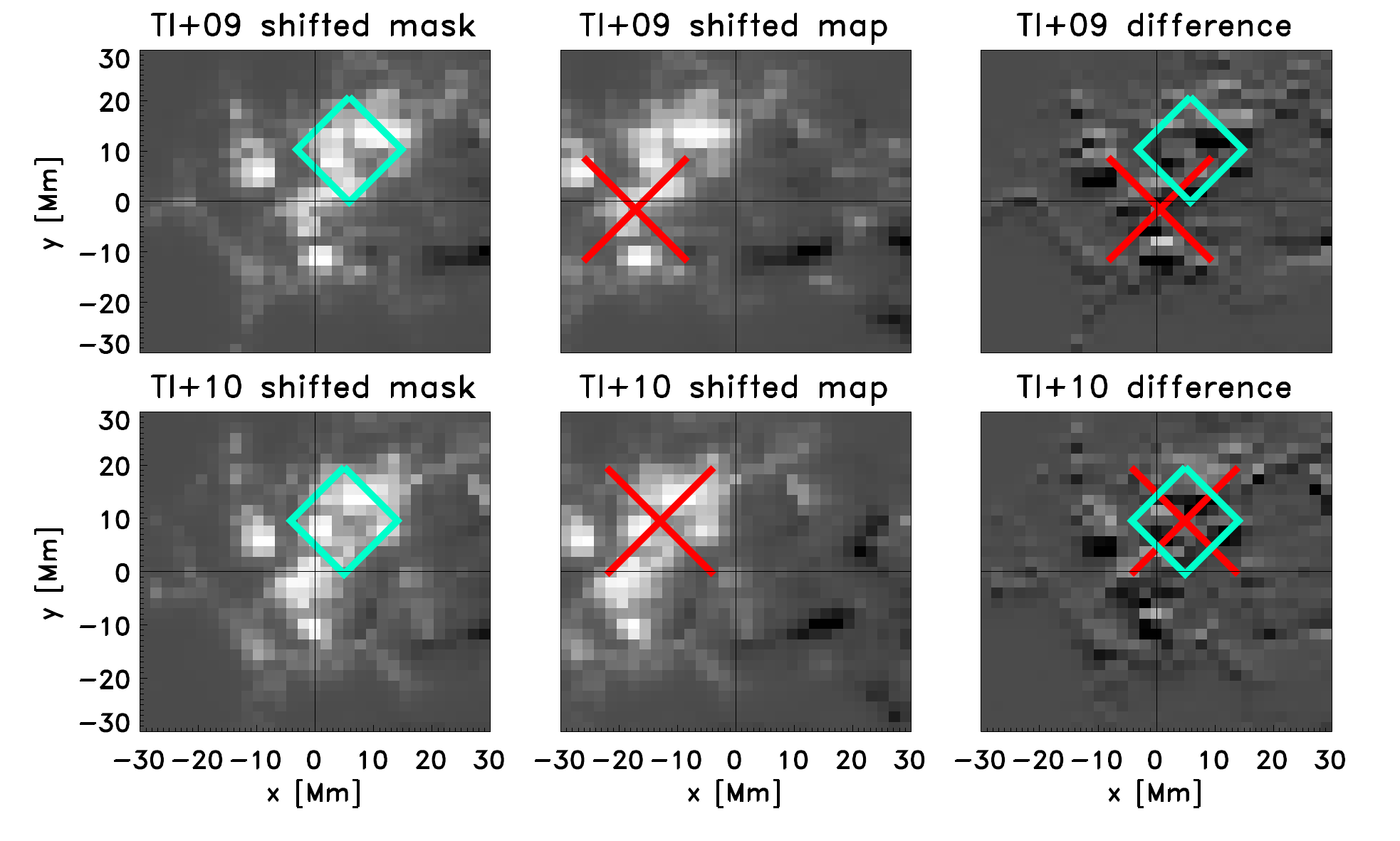}
\caption{
Line-of-sight magnetic field maps and identified locations of the positive (following) polarity at \texttt{TI+09} (top row) and \texttt{TI+10} (bottom row) of example active region 11072. The greyscale is from $-500$~G to 1200~G. The left panel shows the map and location of the polarity (teal diamond) by shifting the search mask.  The middle panel shows the location of the polarity (red cross) after shifting the magnetic field map using bilinear interpolation. 
The size of the symbol is proportional to the radius of the detected feature.
The right panel shows the difference accounting for the integer shift in the maps.  The root mean square of the difference map is 147~G. The largest absolute difference  ($>900$~G) occurs in the white region below the red cross in the \texttt{TI+09} difference map.
}
\label{fig:maps1alg2}
\end{figure*}

\begin{figure*}
\includegraphics[width=0.9\textwidth]{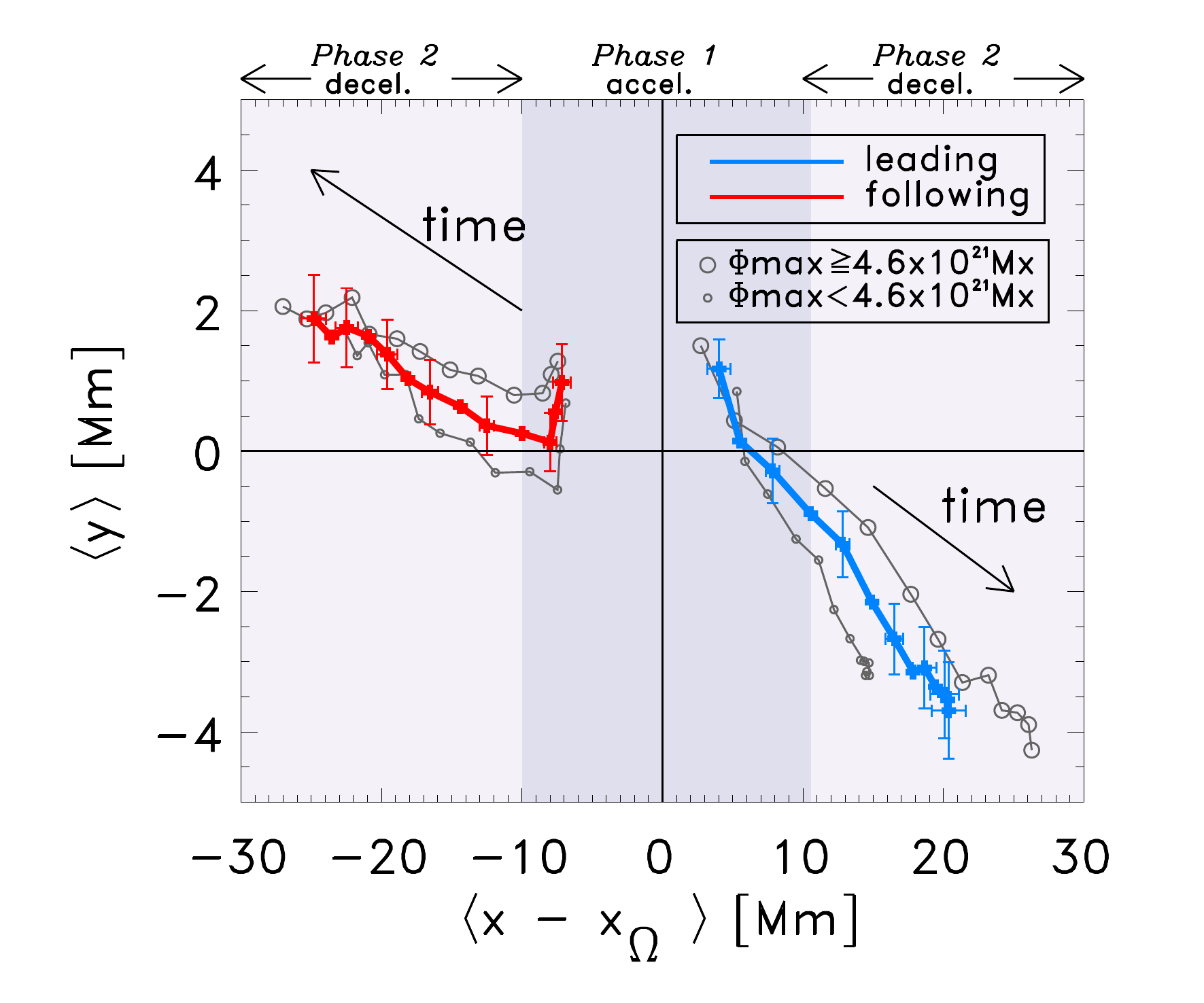}
\caption{
Updated Fig.~2 from \citet{Schunkeretal2019} using the updated algorithm. 
Average over the position of 153  positive (red) and negative (blue) polarities relative to the corrected centre of the map from $\tau=-18.4$~hours (three time intervals, \texttt{TI-03}) before the emergence time, until $\tau=2.1$~days after (\texttt{TI+09}).
The centre of each of the maps were tracked at the Carrington rotation rate \citep{Snodgrass1984}.
We corrected the centre of the map by subtracting the displacement due to difference between the quiet-Sun plasma rotation rate $x_\Omega = R_\odot \Omega(\lambda) \cos(\lambda) \Delta \tau$, where $\lambda$ is the latitude of the centre of the Postel projected map  \cite[see Table~A.1 in ][]{Schunkeretal2016}.
The  blue and red curves cover the time intervals from \texttt{TI-03} to \texttt{TI+09}.
The grey lines with large (small) circles shows the motion of the polarities belonging to regions with maximum flux higher (lower) than the median flux.
The shaded regions indicate Phase~1, when the separation speed between the polarities increases, and Phase~2 when the separation speed decreases \citep[see Fig.~4, ][]{Schunkeretal2019}.
}
\label{fig:mapalg2}
\end{figure*}

Figure~\ref{fig:posalg2} shows an example of the location of the following and leading polarities for AR~11072. This example shows the differences in the location of the leading and following polarity computed by either shifting the magnetic field maps or the search mask, which can be significant for an individual active region. 
Figure~\ref{fig:maps1alg2} shows the maps for \texttt{TI+09} where the bilinear interpolation of the magnetic field map can cause a significant difference in the identified location of the following polarity, and \texttt{TI+10} where it does not. The largest difference is where the bilinear interpolation has introduced values a factor of five larger ($>900$~G, in the white region below the red cross in the \texttt{TI+09} difference map). This moves the centre of gravity of a 25~pixel diameter feature more in this direction. In the case of  \texttt{TI+10}, the interpolator does not introduce such large differences. 
However, in Fig.~\ref{fig:mapalg2}, which is an updated version of Fig.~2, \citet{Schunkeretal2019} shows that there are no significant differences in the {average} polarity positions.

\clearpage 

\section{Polarity centres identified in high-resolution maps}\label{app:hires}

For helioseismology purposes it is sufficient to have a coarser resolution (about 1.4~Mm per pixel) than nominally observed by HMI (about 0.35~Mm per pixel) since the waves are not sensitive to these scales. However, when measuring the location of the polarities, it could be the case that a higher resolution is required for a more precise result.

\begin{figure*}
\includegraphics[width=0.9\textwidth]{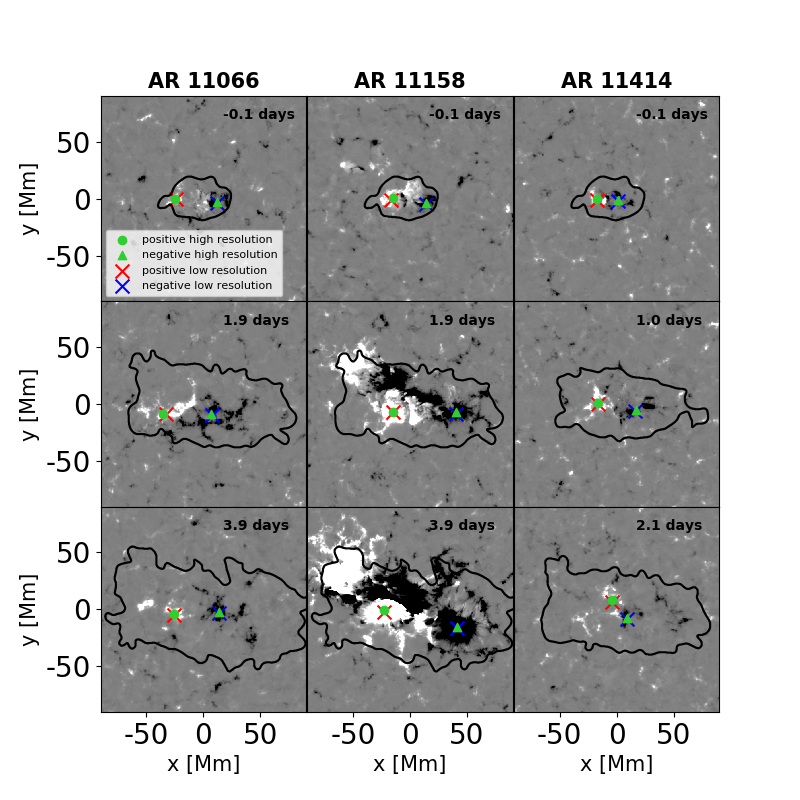}
\caption{
Similar to Fig.~1 in \citet{Schunkeretal2019}.
High-resolution (0.35~Mm per pixel) time-averaged line-of-sight magnetogram maps of AR~11066 (left), AR~11158 (middle), and AR~11414 (right). 
These high-resolution maps can be directly compared with the low-resolution maps in Fig.~1 in \citet{Schunkeretal2019}.
The grey scale is $\pm 1000$~G.
The black contour indicates the search area to identify the polarities in the high-resolution maps. 
The green triangle (circle) shows the position of the negative (positive) polarity identified using the high-resolution maps.
The position of the negative (blue cross) and positive (red cross) polarities computed from the low-resolution (1.4~Mm per pixel) time-averaged line-of-sight magnetograms are shown for comparison. 
}
\label{fig:hiresloc}
\end{figure*}

We repeated our analysis using high-resolution time-averaged line-of-sight magnetic field maps at 0.35~Mm per pixel and shifting the search mask.  The high-resolution maps show more structure in the polarities than the low-resolution maps, particularly for the more dispersed following polarity (see Fig.~\ref{fig:hiresloc}). This makes identifying the primary polarity more difficult at later times, and so we retain the threshold parameters used in the low resolution case, in particular we still search for features with a diameter of 35~Mm. The search area (see black contour in Fig.~\ref{fig:hiresloc}) is based on the average of 78 emerging active regions (listed below), and so it is similar but not identical to the search area used in the main analysis of the paper and  \citet{Schunkeretal2019}. As an example, we show the position of the polarities and the tilt angle for both low- and high-resolution maps of some example active regions in Fig.~\ref{fig:hiresloc}.

\begin{figure*}
\includegraphics[width=0.9\textwidth]{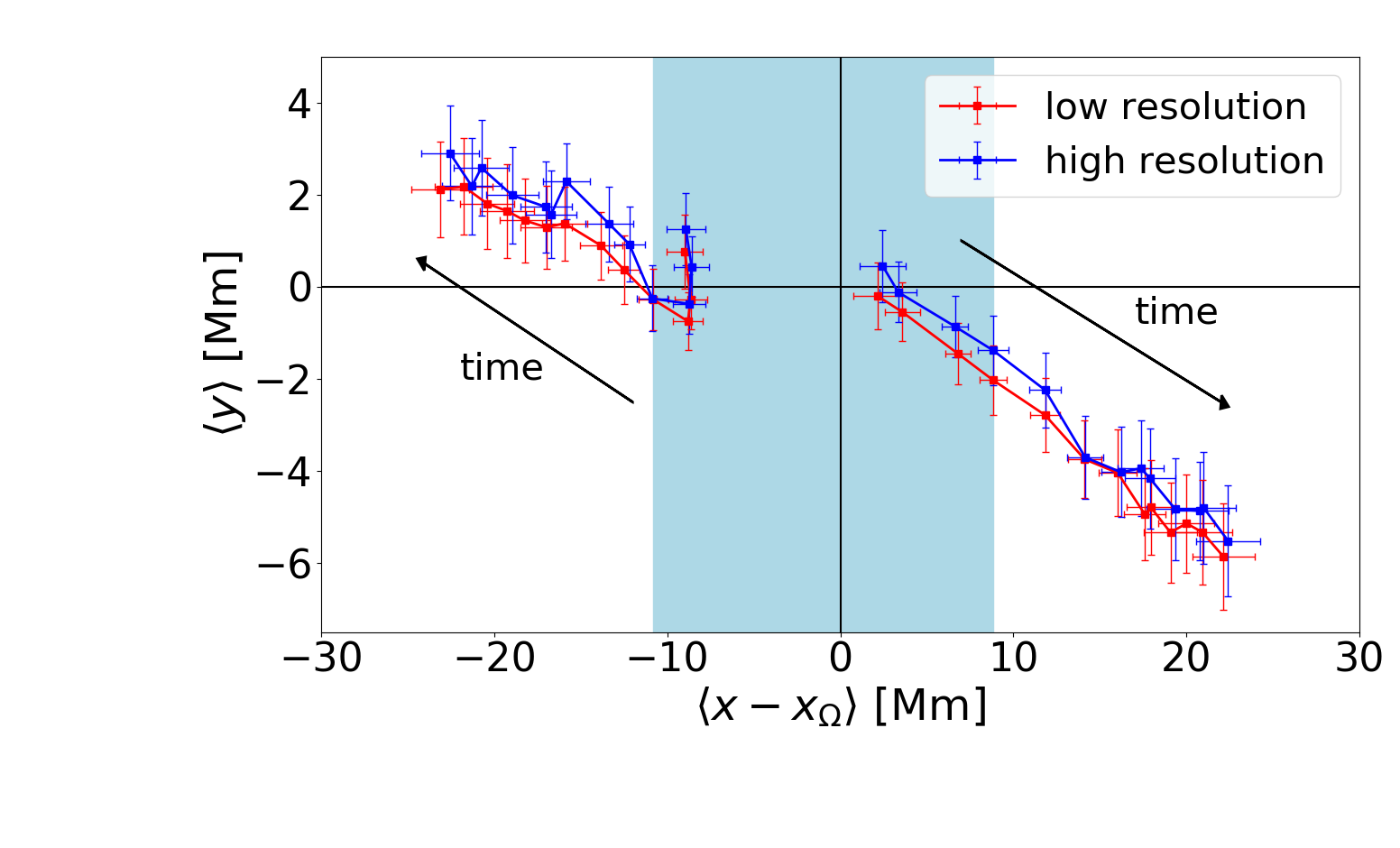}
\caption{
Average position of the polarities from  $\tau=-18.4$~h (three time intervals, \texttt{TI-03}) before the emergence time, until $\tau=2.1$~days after (\texttt{TI+09}) increasing in distance away from the centre. This figure is similar to Fig.~2 \citet{Schunkeretal2019}, except that only 78 of the emerging active regions have been used in both the low-  and high-resolution cases (red and blue, respectively). Differences between the high- and low-resolution cases do not change our previous science conclusions or about the onset of Joy's law. The blue shaded region indicates Phase~1 of the emergence process \citep{Schunkeretal2019} when the polarity separation speed is increasing.
}
\label{fig:hiresbipos}
\end{figure*}

We computed the average position of 78~active regions as a representative subset in the high-resolution maps (Fig.~\ref{fig:hiresbipos}). There is no significant difference between the average position of the polarities in the high- and low-resolution maps of the same 78~active regions.
For our purposes of identifying the location of the primary leading and following polarities, the lower resolution maps suffice.

The subset of 78 emerging active regions used to compare the locations of the polarities in the high- and low-resolution time-averaged line-of-sight magnetogram maps are the following:
11066, 11070, 11072, 11074, 11075, 11076, 11079, 11080, 11081, 11086, 11088, 11098, 11103, 11105, 11114, 11116, 11122, 11130, 11132, 11136, 11137, 11138, 11141, 11142, 11143, 11145, 11146, 11148, 11152, 11154, 11156, 11157, 11158, 11159, 11167, 11174, 11182,
11290, 11291, 11294, 11297, 11300, 11304, 11310, 11311, 11318, 11322, 11326, 11327, 11331, 11334, 11370, 11381, 11385, 11396, 11397, 11400,
11404, 11406, 11414, 11416, 11431, 11437, 11446, 11449, 11450, 11456, 11466, 11472, 11497, 11500, 11510, 11511, 11523, 11531, 11547, 11549,
11551.

\clearpage
\section{Modelling the latitudinal separation between polarities during emergence}\label{app:latsep}

In this section we  explore the change in tilt angle in relation to the Coriolis force, separation speed, and lifetime of the active regions in the SDO/HEARs database.

We  explore three models to describe the north-south displacement:
\begin{enumerate}
\item Constant tilt angle: It has been suggested that the flux tube arrives at the surface already tilted satisfying Joy's law \citep[e.g.][]{Weberetal2013}, but distorted by convection. Here we  test if it is statistically possible that the regions have a constant tilt angle.
We model the north-south displacement due to a constant tilt angle at the time of emergence, as  $\delta y(\tau) = - \tan\left[ \gamma(\tau=0)\right] \delta x(\tau)$.
\item Coriolis effect: Joy's law is a function of latitude, reminiscent of the Coriolis force. \citet{Howard1994}  showed that  bipolar magnetic regions (identified in white light images) that move further apart or closer together change tilt angle in the sense expected from  the Coriolis force acting on this change in separation.
This study only considered day-to-day changes in separation and tilt, and not the evolutionary stage of the bipolar magnetic regions. Here we model the north-south displacement of the polarities due to the Coriolis force given the east-west separation speed and the surface latitudinal differential rotation.
The Coriolis acceleration in the north-south direction, $\delta \ddot{y}_C (\tau)$, acting on a velocity in the east-west direction, $\delta \dt{x}(\tau)$,  in a coordinate system rotating at the local  rotation rate, $\Omega$, is
 \begin{equation}
 \delta \ddot{y}_C (\tau) = -2~\Omega \, \sin \lambda \, \delta\dt{x}(\tau), 
 \end{equation}  
where $\lambda$ is the latitude  and $\tau$ is time. 
The displacement in the north-south direction is related to the velocity and acceleration, $a$ as
\begin{eqnarray}
\delta y(\tau) &=& \int_{\tau_0}^\tau \delta \dt{y}_C (\tau^\prime) \mathrm{d} \tau^\prime + \delta y({\tau_0}) \nonumber \\
&=& \int_{\tau_0}^\tau \left( \int_{\tau_0}^{\tau^\prime} \delta \ddot{y}_C (\tau^{\prime \prime}) \mathrm{d} \tau^{\prime \prime}  + \delta \dt{y}({\tau_0}) \right) \mathrm{d} \tau^\prime + \delta y ({\tau_0}) \nonumber \\
&=& \int_{\tau_0}^\tau \left( \int_{\tau_0}^{\tau^\prime} -2 \Omega \sin \lambda \, \delta\dt{x} (\tau^{\prime \prime}) \mathrm{d} \tau^{\prime \prime} + \delta \dt{y}({\tau_0})\right) \mathrm{d} \tau^\prime
+ \delta y({\tau_0}) \nonumber \\
&=& \int_{\tau_0}^\tau \left( -2 \Omega \sin \lambda \left[ \delta x (\tau^{\prime\prime}) \right]_{\tau_0}^{\tau^\prime} + \delta \dt{y}({\tau_0})\right) \mathrm{d} \tau^\prime + \delta y({\tau_0}) \nonumber \\
&=&-2 \Omega \sin \lambda \left( \int_{\tau_0}^\tau  \delta x(\tau^\prime) \mathrm{d} \tau^\prime  - \delta x ({\tau_0})[\tau - {\tau_0}] \right) \nonumber \\
&+&\delta \dt{y}({\tau_0}) [\tau - {\tau_0}] + \delta y ({\tau_0}) \label{eqn:coriolismodel}
.\end{eqnarray}
In this derivation we assumed that any changes in $\lambda$ in time have a small effect on $\sin \lambda$ and $\Omega (\lambda)$. 
We set ${\tau_0}=0$ and refer to the three terms in this equation separately as 
the Coriolis component of displacement $\delta y_C=-2 \Omega \sin \lambda \left( \int_0^\tau  \delta x(\tau^\prime) \mathrm{d} \tau^\prime  - \delta x (0)\tau \right)$, 
the initial north-south velocity component of displacement, $\delta y_P=\delta \dt{y}(0)\tau$, 
and the initial displacement component, $\delta y_0=\delta y (0)$.
\item Initial north-south velocity component of displacement: Removing the Coriolis effect from Eq.~\ref{eqn:coriolismodel} leaves $\delta y(\tau) = \delta y_P + \delta y_0$.
\end{enumerate}

\begin{figure*}
\includegraphics[width=0.8\textwidth]{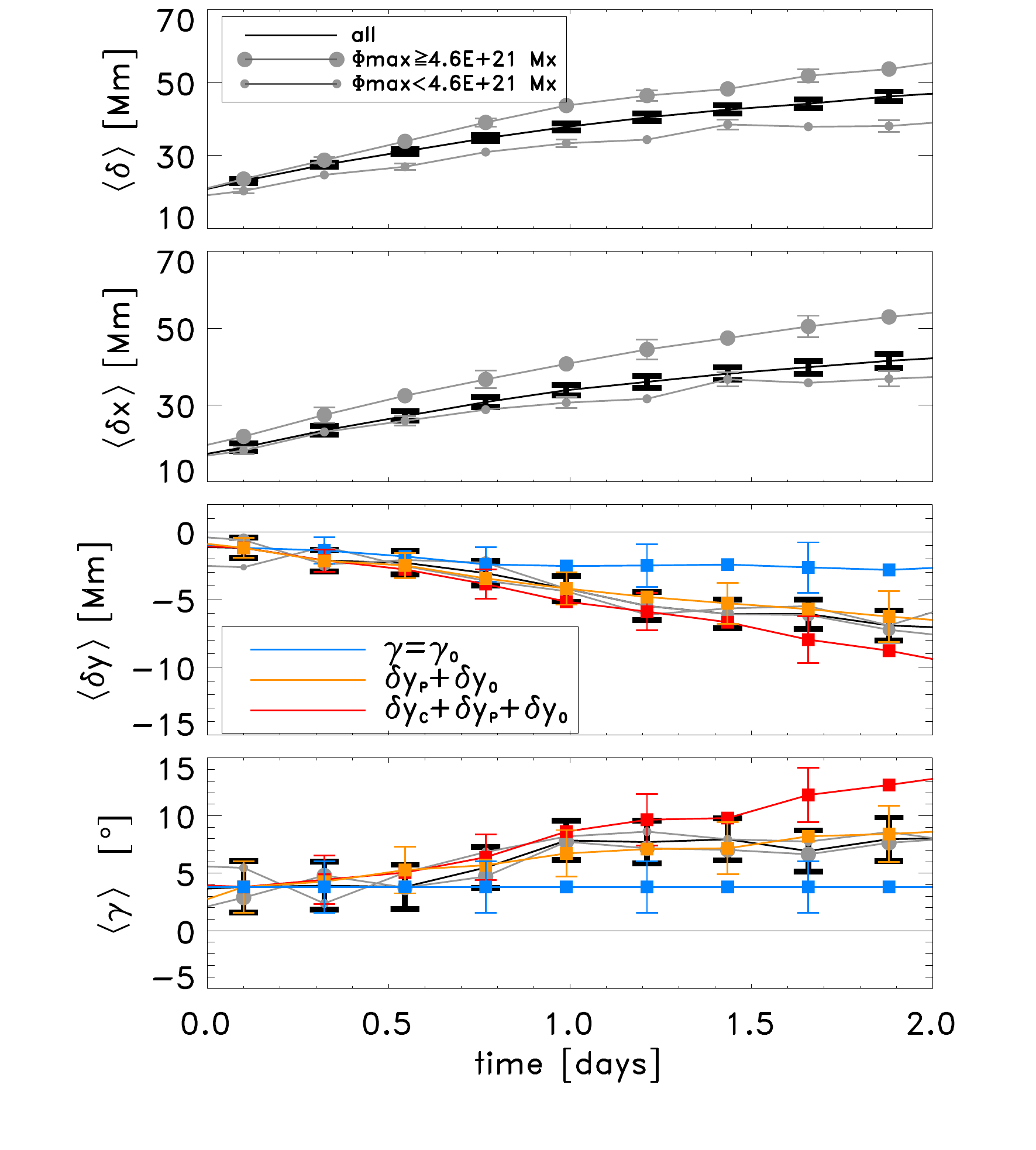}
\caption{
Median separation (top three panels) and tilt angle (bottom panel) of 95 EARs with valid measurements of their location at all times from $\tau=0.1$ to $\tau=2.1$~days  (black curve). This is less than the 153 active regions  used in the body of the paper because in this case    the EAR is required to have a valid measurement at all time intervals shown.
This does not change the results significantly from using all EARs with valid measurements at each time.
High- and low-flux observation samples are indicated by the size of the grey circles.
In the bottom two panels the coloured curves show the different models of displacement and tilt angle, and  grey represents the observations.  
}
\label{fig:distmodels}
\end{figure*}

First we compare the models to the observations as a function of time in the first two days after  emergence. We have selected only the EARs which have valid measurements in each time interval from $\tau=0.1~\mathrm{days}$ to $\tau=2.1~\mathrm{days}$ \cite[see Table~B.1 in ][]{Schunkeretal2019}.
Fig.~\ref{fig:distmodels} shows that the constant tilt angle model can be excluded: the displacement in the north-south direction would not change significantly given the change in separation in the east-west direction.
The north-south displacement due to the constant, initial north-south velocity component of displacement agrees with the measured displacement best, and the addition of the Coriolis effect acting on the east west separation speed is relatively small.

\begin{figure*}
\includegraphics[width=0.9\textwidth]{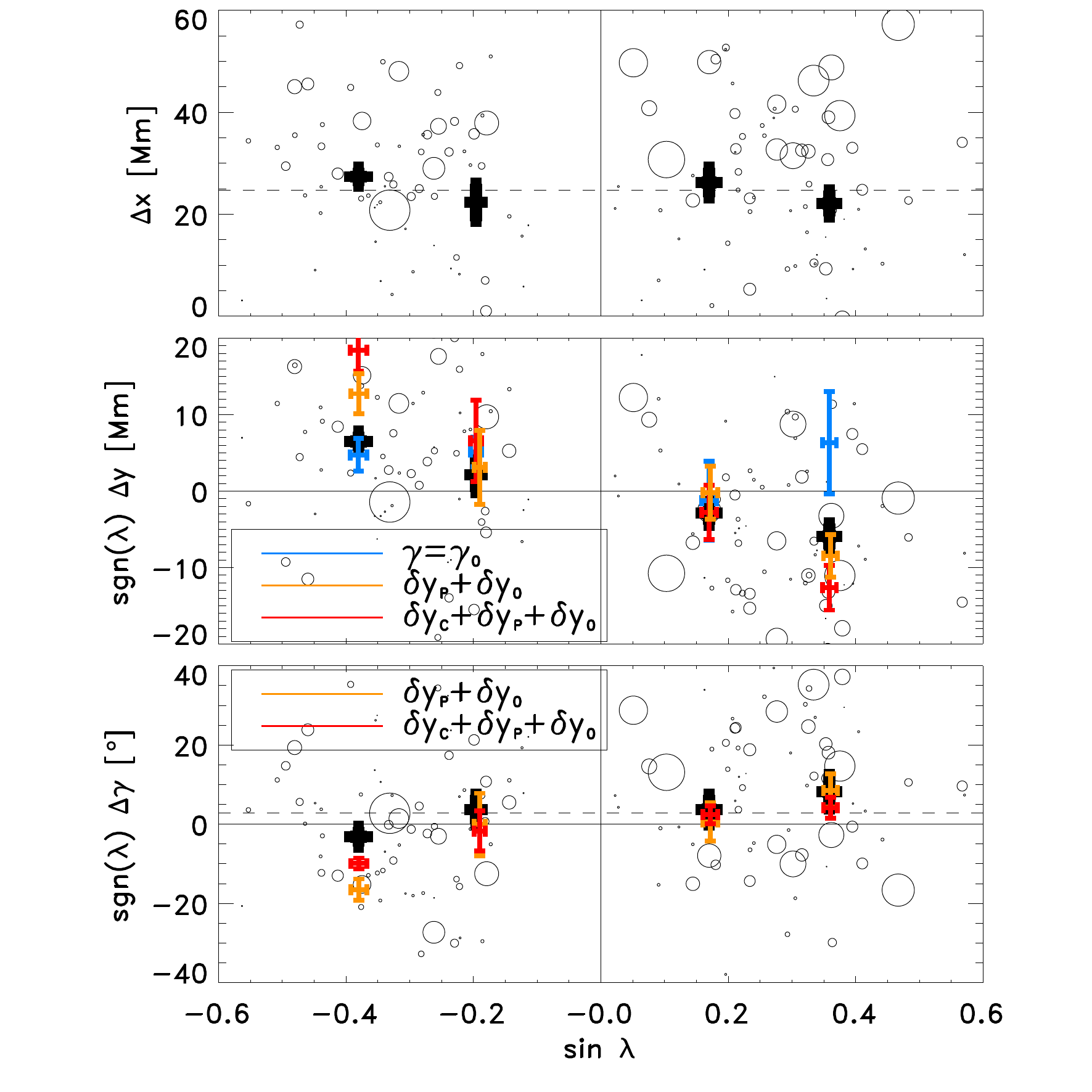}
\caption{
Change in east-west separation, $\Delta x$, (top panel); north-south separation, $\Delta y$, (second panel); and change in tilt angle, $\Delta \gamma$ (bottom panel) in the first two days after emergence as a function of latitude, $\lambda$. 
The size of the circle is proportional to the maximum flux of the active region.
The thick black points with error bars show the averages for active regions in different latitude ranges. 
The coloured points with error bars show latitudinal averages (between 0$^\circ$ and 15$^\circ$ ($\sin \lambda = 0.26$) and 15$^\circ$ and 40$^\circ$ ($\sin \lambda = 0.64$) in the northern hemisphere, and the equivalent in the southern hemisphere) for the modelled north-south separation and change in tilt angle for the constant tilt angle model in blue, the Coriolis effect in red, and the initial north-south velocity of the polarities in yellow.
 }
\label{fig:sinelat}
\end{figure*}

We then compare the models to the observations as a function of latitude.
We measure the change over the first two days after emergence in the separation between the polarities, 
$\Delta x = \delta x (\tau=2.1~\mathrm{days}) - \delta x (\tau=0.1~\mathrm{days})$ and 
$\Delta y = \delta y (\tau=2.1~\mathrm{days}) - \delta y (\tau=0.1~\mathrm{days})$, and 
the change in tilt angle, 
$\Delta \gamma = \gamma (\tau=2.1~\mathrm{days}) - \gamma (\tau=0.1~\mathrm{days})$.
The time interval is given in units of days  \cite[see Table~B.1 in ][]{Schunkeretal2019}. 
Figure~\ref{fig:sinelat} shows the change in displacement and tilt angle over two days as a function of latitude. 
The separation in the east-west direction is not dependent on latitude, however, the separation in the north-south direction and the tilt angle is, showing that the tilt angle comes predominantly from the north-south motion.
Again, we see that the constant tilt angle model cannot explain the north-south displacement.

\end{document}